\documentstyle[aps,epsf]{revtex}
\begin{document}

\twocolumn [
\hsize\textwidth\columnwidth\hsize\csname@twocolumnfalse\endcsname

\title{ Landau damping and 
the echo effect in a confined  Bose-Einstein condensate }
\author{A. B. Kuklov}
\address{ Department of Applied Sciences,
The College of  Staten Island, CUNY,
     Staten Island, NY 10314}

\maketitle
\begin{abstract}
Low energy collective mode of a confined Bose-Einstein
condensate should demonstrate the echo effect in the 
regime of Landau damping.
 This echo is  a signature of  reversible nature of 
 Landau damping. 
General expression for the echo profile is derived in the limit 
of small amplitudes of the external pulses. 
Several universal features of the echo are found. 
The existence of echo in other cases of reversible
damping -- Fano effect and Caldeira-Leggett model --
is emphasized. 
It is suggested to test  reversible
nature of the damping in the atomic traps by conducting the echo
experiment.
\\

\end{abstract}
\vskip0.5 cm
]

Recent achivements in trapping and cooling alkaline gases  
 \cite{BEC} have raised 
 strong interest to various aspects of many body collective
behavior. In many respects, properties of a confined atomic
cloud forming Bose-Einstein condensate are unique. For example, 
low energy collective modes of  Bose-Einstein 
condensate confined at nearly absolute
zero temperature demonstrate damping whose rate is several times larger
than that in the thermal cloud \cite{DAMP}. Now it is commonly accepted
that the damping in a confined condensate should be Landau damping (LD) typical
when collisions between quasiparticles are rare events \cite{PIT,LU,SHLAP,GIO}.
It has been especially emphasized in Ref.\cite{SHLAP} that the LD occurs
in anisotropic traps where the spectrum exhibits random character. 
 
Landau damping (LD) occurs due to a resonance interaction between a 
collective mode and quasiparticle excitations. This damping 
does not lead to a thermalization of the quasiparticle distribution.
Therefore, it is essentially a reversible phenomenon of dephasing
of the collective mode due to mixing with the quasiparticle continuum
in a sense of the Fano effect \cite{FANO} and Caldeira-Leggett model
\cite{CALD}. 

A main test of the reversibility of the damping
is the echo effect. In the conventional phonon echo model
(see in Ref. \cite{MASON}), a phonon mode is characterized
by an inhomogeneous broadening. As a result, the amplitude
of the oscillations initiated at the time moment 
$t=0$ decays in time. Second short pulse imposed
on this mode at $t=\tau$ creates a partial time reversal of the dynamics
causing the echo at the time moment $t=2\tau$. 
This type of echo is called $e_2$ \cite{MASON}. 
The phonon echo is a single mode effect whose essential ingridient
is a non-linearity of the mode \cite{MASON}.  

The case under consideration is the multimode effect. 
The essence of the echo effect 
considered here can be briefly outlined
as follows: externally excited collective mode interacts resonantly with other
degrees of freedom -- quasiparticles constituting a thermal component.
As a result, the amplitude of the mode decays
in time. A second external pulse 
partly reverses in time the evolution of the whole system. Hence, a 
part of the energy returns back to 
the collective mode at the time moment which approximately equals to twice
 the time interval between the two external pulses.
In accordance with the classification \cite{MASON}, this
type of echo can be called $e_2$. As it will be shown below,
a crucial element for this echo is a parametrical excitation 
of the thermal component. In its nature the echo mechanism
described above is close to the echo   
in classical uniform 
 collisionless plasma \cite{PLASMA}, where it occurs 
 at the wave vector
equal to the difference between the wave vectors of the imposed pulses
\cite{PLASMA}. Accordingly, the delay between the echo and the second
pulse can be controlled externally. The amplitude of the plasma echo
\cite{PLASMA}
turns out to be of the first order with respect to the amplitudes
of the both pulses. These properties are due to two circumstances: 1)
the uniformity of the plasma giving rise to the momentum conservation,
and 2) the quadratic non-linearity of the kinetic equation \cite{PLASMA}.   
In a non-uniform confined condensate, the condition 1)
 does not exist. Therefore,
the plasma type echo \cite{PLASMA} should not be a universal feature
of such a condensate. 

In Ref.\cite{ECHO} the $e_2$-echo effect has been predicted for 
 isotropic condensate, where no LD is expected to occur \cite{SHLAP}
and where a major mechanism of damping should be thermal
dephasing \cite{ECHO}.
In this paper the echo in the presence of the LD in a confined condensate
 will be considered.
It will be shown that the $e_2$-echo occurs in this case. Its
amplitude turns out to be linear in the initial amplitude of the 
collective mode and is quadratic in the amplitude of the second pulse.
The time profile of the echo response demonstrates a specific double
peak structure.

In order to describe the LD and the echo effect in the case of
a many boson system  demonstrating the phenomenon of Bose-Eisntein
condensation in a confined geometry, the standard form 

\begin{eqnarray}
 H=\int d{\bf r}  \Psi^{\dagger}[H_1 +{g \over 2}\Psi^{\dagger}\Psi ]\Psi,
\label{eq:H} \\
\displaystyle H_1= - {\hbar^2 \over 2m}\nabla^2 + V_{ex}-\mu
\label{eq:H1}
\end{eqnarray}
\noindent
of the Hamiltonian is employed. 
Here the Bose operators $\Psi , \Psi^{\dagger}$
obey the usual Bose commutation rule;
 $\mu$ is the
 chemical potential; $V_{ex}$ denotes the trapping potential;
$g$ stands for the interaction constant; and $m$  is
atomic mass. In what follows the atomic units ($\hbar =1$) will be employed.

Below we will closely follow the approach employed in Ref.\cite{GIO}.
Accordingly, two types of averaging $\langle ... \rangle$ and 
$\langle ... \rangle_{eq}$ are introduced. The first one is performed
over the initial state, while the second is the equilibrium thermal
averaging. 
Thus the Bose field $\Psi$,
the excited condensate wave function
$\Phi$ and the equilibrium condensate wave function $\Phi_0$
can be represented as 

\begin{equation}
\Psi = \Phi + \tilde{\Psi}, \, \Phi = \langle \Psi \rangle , 
\, \Phi_0=\langle \Psi \rangle_{eq}.
\label{eq:Phi}
\end{equation}
\noindent
 Here $\tilde{\Psi}$ stands for the non-condensate part.
In order to simplify the following consideration, the Hartree approximation
for the thermal component described by $\tilde{\Psi}$ will be employed.
 As it is well known, this
approximation is valid at high temperatures \cite{KONDOR}. 
An extension to general case will be considered elsewhere. 
Consequently, we ignore the anomalous mean
 $\langle  \tilde{\Psi}  \tilde{\Psi}\rangle$ and retain only 
the quantity $
\tilde{n}= \langle \tilde{\Psi}^{\dagger}({\bf x})
 \tilde{\Psi}({\bf x})\rangle$ 
in the equation for the condensate wave function $\Phi$.

Equations for $\Phi$, $\tilde{\Psi}$ have be obtained in Ref. \cite{GIO}.
Here a Hartree limit of these will be utilized. For this purpose the Heisenberg
equation $i\hbar \dot{\Psi}=[\Psi , \,H]$ is averaged over $\langle ...\rangle$
and the three operator mean $\langle \Psi^{\dagger}\Psi \Psi \rangle$
 is factorized after the substitution of
the representation (\ref{eq:Phi}) has been made \cite{GIO}. Then, the
factorization $\tilde{\Psi}^{\dagger}\tilde{\Psi}\tilde{\Psi}
\to 2\langle \tilde{\Psi}^{\dagger}\tilde{\Psi}\rangle \tilde{\Psi}+
\langle \tilde{\Psi}\tilde{\Psi}\rangle \tilde{\Psi}^{\dagger}$ 
\cite{GIO} is made in the equation for $\tilde{\Psi}$. 
Finally, in the Hartree limit these equations are

\begin{eqnarray}
i\hbar \dot{\Phi}= [H_1 + g|\Phi|^2 + 2g\tilde{n}]\Phi ,
\label{eq:GP}\\
i\hbar \dot{\tilde{\Psi}}= [H_1 + 2g(|\Phi|^2 +\tilde{n})]\tilde{\Psi} .
\label{eq:exc}
\end{eqnarray}
Eqs.(\ref{eq:GP}), (\ref{eq:exc}) together with the definition of $\tilde{n}$
determine completely the mean field dynamics
of the condensate and the thermal component in the 
Hartree limit ($T >> \mu $) in the collisionless regime.
In this regime the only mechanism of damping is the 
LD which is rather a reversible dephasing of collective oscillations 
than their irreversible thermalization. 

It is enough to consider linearized dynamics
of the quantities $
\Phi'=\Phi - \Phi_0,\, n'=\tilde{n} - \tilde{n}^0,
$
describing small deviations from the equilibrium, where the notation
$ \tilde{n}^0=\langle \tilde{\Psi}^{\dagger} \tilde{\Psi}\rangle_{eq}$
is employed \cite{GIO}.
In the Bogolubov
representation for the condensate part and the Hartree approximation 
for the thermal component 

\begin{eqnarray}
\Phi'=\sum_n (u_na_n + v_n^*a^*_n),
\label{eq:Bog1}\\
\tilde{\Psi}=\sum_n u_nb_n ,
\label{eq:Bog2}
\end{eqnarray}
\noindent
 $a_n$, $a^*_n$ are classical numbers and
the operators $b_n$ and $b^{\dagger}_n$ destroys and creates, respectively, 
 thermal particle on the level
with the energy $\varepsilon_n$. The values $u_n, v_n$ form 
the eigenvector of the Bogolubov system

\begin{eqnarray}
\varepsilon_nu_n=H'_1u_n +g\Phi_0^2v_n, \\
-\varepsilon _nv_n=H'_1v_n + g\Phi_0^{*2}u_n, \\
H'_1=H_1 + 2g(|\Phi_c|^2+\tilde{n}^0),
\label{eq:bog} 
\end{eqnarray}
\noindent
and obey the orthogonality condition

\begin{equation}
\int d{\bf x}[ u^*_m({\bf x}) u_n({\bf x})-
 v^*_m({\bf x}) v_n({\bf x})]=\delta_{mn}.
\label{eq:orto}
\end{equation}
For the thermal part in the Hartree approximation,
one should set the "hole" part $\sim v_n$ to zero.
Then employing Eqs.(\ref{eq:GP}), (\ref{eq:orto}), one obtains

\begin{equation}
\displaystyle i\dot{a}_m=\varepsilon_ma_m + 2g\sum_{kl}
A_{mkl}f_{kl}
\label{eq:a_m}
\end{equation}
\noindent
where the notations 

\begin{eqnarray}
A_{mkl}=\int d{\bf x} \Phi_0(u^*_m+v^*_m)u_k^*u_l,
 \label{eq:A}\\
f_{kl}=\langle b^{\dagger}_kb_l \rangle - f^{(0)}_{kl},
\quad f^{(0)}_{kl}=\langle b^{\dagger}_kb_k \rangle_{eq}=
f^{(0)}_k\delta_{kl},
\label{eq:f_kl}
\end{eqnarray}
\noindent
are introduced, with $f^{(0)}_k=1/(\exp(\varepsilon_k)-1)$ being 
the equilibrium population at the $k$-th level. 
 We
employ Eqs.(\ref{eq:exc}),(\ref{eq:Bog2}), (\ref{eq:orto}), (\ref{eq:f_kl})
and obtain the linearized equation for $f_{kl}$ (see Ref.\cite{GIO}) as

\begin{equation}
\displaystyle i\dot{f}_{kl}=\omega_{kl}f_{kl}+ 2g
(f^{(0)}_k- f^{(0)}_l) \sum_m(A^*_{mkl}a_m + A_{mlk}a^*_m),
\label{eq:f}
\end{equation}
\noindent
where $\omega_{kl}=\varepsilon_l -\varepsilon_k$.

Let us assume that an external resonant drive,
imposed on the system at times $-\infty < t <0$, has prepared
a state with some $a_1(0)\neq 0$ and the rest $a_n(0)=0,\,n\neq 1$. 
We also assume that the thermal component is not affected
by this resonant drive, so that $f_{kl}(0)=0$.
The evolution between the time 
moments $t=0$ and some $t=\tau$ is characterized by Landau damping
due to the terms $\sim f_{kl}$ in Eq.(\ref{eq:a_m}), so that
the amplitude of the oscillations set by the drive at $t=0$
will decay exponentially at later times. 
The evolution 
of the excited mode $a_1(t)$ can be obtained 
from Eqs.(\ref{eq:a_m}),(\ref{eq:f}). In the lowest order 
one should ignore all the low energy modes but $a_1(t)$. Then the
rate of the exponential damping follows as \cite{GIO}
$\gamma_L=\gamma_L(\varepsilon_1)$ where

\begin{equation}
\displaystyle  \gamma_L (\omega ) = 4\pi g^2 \sum_{kl}( f^{(0)}_k- f^{(0)}_l) 
\delta (\omega - \varepsilon_l + \varepsilon_k)|A_{mkl}|^2.
\label{eq:gamma}
\end{equation}

It is worth noting that Eqs.(\ref{eq:a_m}), (\ref{eq:f}) can be
interpreted in a sense of the Fano effect \cite{FANO}.
Indeed, the state $\varepsilon_1$ can be considered
as a discrete state mixed with a quasi-continuum
of the pair excitations
characterized the  spectrum $\omega_{kl}$. As a result, the discrete
state autoionizes with the mean life time $1/\gamma_L$ \cite{FANO}.

Now let us consider the implications of imposing a short 
external pulse at the time moment $t=\tau $ since the state
with $a_1(0)\neq 0$ was created. Let us assume
that $\tau > 1/\gamma_L$ so that the solution
$a_1(t)$ has apparently dissipated at the moment of the pulse.
This pulse can be introduced 
by means of  varying 
the external potential $V_{ex}$, 
so that $V_{ex} $ is replaced by $ V_{ex} + \delta V_{ex}(t)$
 in $H_1$ (\ref{eq:H1}) with

\begin{equation}
\delta V_{ex}= V_2(r)\delta(t-\tau),
\label{eq:drive}
\end{equation}
\noindent
where $V_{2}(r)$ denotes some time independent 
coordinate function describing
strength of the external pulse.

 Below it will be shown
that this pulse will cause the echo -- a  stimulated revival
of the mode $a_1(t)$ in the form of a double peak centered close
to the time moment $t=2\tau$.   
Note that, in contrast with a standard plasma echo \cite{PLASMA}
which essentially requires a non-linearity,
 the echo in the considered case 
is expected to exist in the linear approximation
given by Eqs.(\ref{eq:a_m}),(\ref{eq:f}), if the parametrical
 excitation of the thermal
component by the second pulse is taken into account.
This excitation produces a  
 change of $f_{kl}(t)$ while $t$ crosses
the moment $t=\tau$, so that $ f_{kl}(\tau + \epsilon)$
acquires an admixture of its complex conjugate
 $ f_{kl}(\tau)^*= f_{lk}(\tau)$.
This fact constitutes a partial time inversion of the original
evolution of the solution $a_1(t)$. 
Considering
the solution of Eq.(\ref{eq:exc}), where $H_1$
has been modified by (\ref{eq:drive}),  in the limit $V_2(r)\to 0$
and ignoring the non-linearity of Eq.(\ref{eq:exc})
during the action
of the $\delta$-pulse (\ref{eq:drive}),
one finds $\tilde{\Psi}(\tau + \epsilon)= S[V_2]
\tilde{\Psi}(\tau)$.
Here  $S[V_2]\approx 1-iV_2$ stands for the evolution operator
transforming the solution from the time just before
the pulse to the time just after the pulse. Employing the
 orthogonality condition (\ref{eq:orto}) and Eq.(\ref{eq:Bog2})
in the Hartree approximation, one can find $b_m(\tau + \epsilon)=
\sum_n S_{mn}b_n(\tau)$, and finally Eq.(\ref{eq:f_kl}) yields

\begin{eqnarray}
\displaystyle f_{mn}(\tau + \epsilon)=
\sum_{ks}S^*_{mk}S_{ns}f_{ks}(\tau),
\label{eq:fjump}\\
S_{mn}=\int d{\bf x}u^*_mS[V_2]u_n .
\label{eq:S}
\end{eqnarray}
\noindent
The term responsible for the echo corresponds to $k=n,\,s=m$ in the
sum (\ref{eq:fjump}). Therefore, in the lowest order with respect
to $V_2$ one finds 

\begin{eqnarray}
\displaystyle f_{mn}(\tau + \epsilon)=
|(V_2)_{mn}|^2f_{nm}(\tau)+ \Sigma ',
\label{eq:fecho}\\
(V_2)_{mn}=\int d{\bf x}u^*_mV_2u_n .
\label{eq:Vmn}
\end{eqnarray}
\noindent
where $\Sigma '$ represents terms which do not lead to the echo
and which will be omitted in the following analysis.

It is convenient to introduce a quantity $A(t)=a_1(t)+a^*_1(t)$.
Given the symmetry of the original Hamiltonian
with respect to time inversion and the representation (\ref{eq:A})
one finds $ A^*_{1mn}=A_{1mn}=A_{1nm}$. Then Eq.(\ref{eq:a_m})
 yields 
$\dot{A}(t)
=-i\varepsilon_1 (a_1(t) - a^*_1(t))$.  
The echo time profile can be obtained by finding the
Laplace transforms  
of (\ref{eq:a_m}),(\ref{eq:f}) on the time interval $t>\tau$
and expressing them in 
terms of 
 $ f_{mn}(\tau + \epsilon )$ and $A(\tau + \epsilon ),\,
\dot{A}(\tau + \epsilon )$. The echo term resides in the 
part containing $ f_{mn}(\tau + \epsilon )$. Thus, the 
contributions $\sim A(\tau + \epsilon ),\,
\dot{A}(\tau + \epsilon )$ can be omitted. Then, $ f_{mn}(\tau + \epsilon )$
is expressed by means of Eq.(\ref{eq:fecho}). After that
the quantity $f_{mn}(\tau )$ can be obtained as the inverse
transform on the time interval $0<t<\infty$ with $f_{mn}(0)=0$
and $A(0)\neq 0,\, \dot{A}(0) \neq 0$ as though
no pulse at $t=\tau$ is present \cite{CASUAL}. 
In the solution $A(t) $ only that part $A^{ec}(t)$ which 
contains the $e_2$-echo should be retained.  Finally,  
one finds the solution $A^{ec}(t) $ for
 $t>\tau $ as

\begin{equation}\begin{array}{c}
\displaystyle A^{ec}(t)={8\varepsilon^3_1 \over \pi^3}\int 
d\omega \int d\omega '
{\rm e}^{-i\omega(t-\tau) -i\omega '\tau} \\ \\
\displaystyle \int d\varepsilon {\tilde{\gamma}(\varepsilon)\over
(\omega - \varepsilon) (\omega ' + \varepsilon)} \\ \\
\displaystyle  
{\gamma_L (\omega) \gamma_L (\omega ')[i\dot{A}(0) + \omega ' A(0)] \over
[(\omega^2 - \varepsilon^2_1)^2+(2\varepsilon_1 \gamma_L(\omega))^2]
[(\omega'^2 - \varepsilon^2_1)^2+(2\varepsilon_1 \gamma_L(\omega')^2]},
\label{eq:echo1}
\end{array}\end{equation}
\noindent
where the notation

\begin{equation}
\displaystyle \tilde{\gamma} (\varepsilon)= 4\pi g^2 \sum_{mn}
 |(V_2)_{nm}|^2A_{1mn}^{*2}(f^{(0)}_m- f^{(0)}_n)\delta (\varepsilon -
\varepsilon_n +  \varepsilon_m)
\label{eq:tildegamma}
\end{equation}
\noindent 
is employed. In fact, all the information about the pulse
(\ref{eq:drive}) is contained in the quantity $\tilde{\gamma}$
(\ref{eq:tildegamma}). 

The double
singularity $1/(\omega - \varepsilon)(\omega ' + \varepsilon)$ in 
 Eq.(\ref{eq:echo1})
gives rise to the term $-\pi^2\delta (\omega +\omega')\delta(\varepsilon 
-\omega )$ plus regular terms,
in accordance with
Ref.\cite{FANO}.
Omitting these regular terms and performing
trivial integrations over $\varepsilon$ and $\omega'$,
one arrives at Eq.(\ref{eq:echo1})
rewritten as

\begin{equation}\begin{array}{c}
\displaystyle A^{ec}(t)={8\varepsilon^3_1 \over \pi}\int d\omega 
{\rm e}^{-i\omega(t-2\tau)}\\  \\
\displaystyle
{\tilde{\gamma}(\omega)\gamma_L (\omega)^2 [i\dot{A}(0) - \omega  A(0)] \over
[(\omega^2 - \varepsilon^2_1)^2+(2\varepsilon_1 \gamma_L(\omega))^2]^2}.
\label{eq:echo2}
\end{array}\end{equation}
\noindent
This equation is the main result of the present work. First, it indicates that
the echo is of the $e_2$ type \cite{MASON} due to the exponent
$\exp (i\omega(t-2\tau))$. Second, the integrand 
Eq.(\ref{eq:echo2}) has complex poles of the second order. 
 This determines a specific double peak structure of the
echo response. Indeed, setting $A(0)=0,\, \gamma_L(\omega)=\gamma_L,\,
\tilde{\gamma}_L(\omega)=\tilde{\gamma}_L(\varepsilon_1)=\tilde{\gamma}$, 
one finds the integral (\ref{eq:echo2}) as

\begin{equation}
 A^{ec}(t)={2\dot{A}(0) \tilde{\gamma} \over \varepsilon_1}(t-2\tau)
{\rm e}^{-\gamma_L|t-2\tau|}\sin(\varepsilon_1(t-2\tau)),\quad t>\tau.
\label{eq:split}
\end{equation}
\noindent 
Note that the echo becomes maximal at $|t-2\tau|\approx 1/\gamma_L$ 
and reaches the value $A^{max} \sim \dot{A}(0) \tilde{\gamma}/\gamma_L$
which does not depend explicitly  on the interaction constant $g$,
as a comparison of Eqs.(\ref{eq:tildegamma}), (\ref{eq:gamma}) 
indicates. On the other hand,
$A^{max}$ is proportional to the first power of the initial
amplitude of the collective mode $A(t)$ at $t=0$ and is of the second order
with respect to the amplitude of the second pulse $V_2\to 0$.
These features are universal and do not depend on details of the
matrix elements. The only requirement is that the sum (\ref{eq:tildegamma})
is well defined at $\varepsilon = \varepsilon_1$. In anisotropic
traps the spectrum of the pair excitations can be considered as
a continuum \cite{SHLAP}. Therefore, this sum can be replaced by
integration which yields a finite value of $\tilde{\gamma}$.
In contrast, a temperature dependence of $A^{max}$ is sensitive to  
details of the matrix elements in Eqs.(\ref{eq:tildegamma}),
(\ref{eq:gamma}).

 In the presence of irreversible dissipation determined by the higher
correlators and 
characterized by some rate $\gamma_{(irr)}$ the echo amplitude 
is to be exponentially suppressed. It is straightforward to realize
that the maximum echo amplitude acquires an extra factor
$ \exp(-2\tau \gamma_{(irr)})$. Therefore, in order to observe
a distinct echo, the condition $ 1/\gamma_L <\tau <1/\gamma_{(irr)}$
should be fulfilled. 
 
It is worth noting that the structure of the echo response represented
by Eqs.(\ref{eq:echo2}), (\ref{eq:split}) is typical for other
models of reversible damping as well.
Let us discuss this for the Caldeira-Leggett
model \cite{CALD}. In this case some oscillator $Q$ interacts
with a quasi-continuum of the oscillators $X_i$. The $e_2$-echo
discussed above occurs as long as a parametrical excitation of
the quasi-continuum is imposed. Such an excitation can be introduced
by the time-dependent 
energy term $H'(t)=\sum_i\delta \omega^2_i(t)X_i^2/2$, where
$\delta \omega^2_i(t)$ describes the time dependent part of the
frequency of the $i$-th oscillator. If at the moment $t=0$ some
perturbation $Q(0)\neq 0,\, \dot{Q}(0) \neq 0$ was created, it will
decay due to interaction with the bath oscillators $X_i$ \cite{CALD}.
However, if the parametrical drive of the bath frequencies
is imposed in the form
$\delta \omega^2_i(t)=k_i\delta (t-\tau)$, the amplitude
$Q(t)$ will demonstrate a stimulated revival -- the $e_2$-echo. 
The profile of this echo is exactly given by 
Eqs.(\ref{eq:echo2}), (\ref{eq:split}), where $\tilde{\gamma}$ 
(\ref{eq:tildegamma}) is replaced by a structure 
which turns out to be of the first order with respect to $k_i$. 
Accordingly,  the echo amplitude becomes of the first order
with respect to the amplitude of the external drive. A detailed
discussion of the echo effect in the Caldeira-Leggett model
will be given elsewhere. 

In summary, it is shown that the $e_2$-echo can be observed in a confined
Bose-Einstein condensate in the regime of collisionless damping. 
 It has a specific two-peak structure. 
The echo effect described above could be observed in the traps 
employed in Refs.\cite{DAMP}. The $e_2$-echo is typical for the
Caldeira-Leggett model as well.


\begin{references}
\bibitem{CASUAL}
In accordance with casuality, the pulse at $t=\tau$
cannot affect the evolution at $t<\tau$. Therefore
the integration in the Laplace transformation can be carried
out in the region $0<t<\infty$ as though no pulse
occurs.
\end{references}
\end{document}